\RequirePackage[]{graphicx} 

\def\mode{1} 

\if 1\mode
\documentclass[journal]{article}
\fi

\if 2\mode
\documentclass[num-refs]{wiley-article_final}	
\fi

\usepackage{etoolbox}
\newtoggle{ARXIV}

\newtoggle{SUPMATERIAL}
\togglefalse{SUPMATERIAL}

\usepackage[utf8]{inputenc}
\usepackage{amsmath}
\usepackage{bm}
\usepackage{bbold}
\usepackage{authblk}
\usepackage{amssymb}
\usepackage{graphicx}
\usepackage{ifdraft}
\usepackage{marginnote}
\usepackage{pdfpages}
\usepackage{tabularx,booktabs}
\usepackage{ifpdf}
\usepackage{float}
\usepackage{authblk}
\usepackage[font=small]{caption}
\usepackage[utf8]{inputenc}
\usepackage[english]{babel}
\ifpdf
\usepackage[]{hyperref}
\else
\usepackage[dvipdfm]{hyperref}
\fi
\usepackage{tikz}
\usepackage[capitalise]{cleveref}

\usepackage{blindtext}
\usepackage{amsmath,bm, amsfonts,amssymb,amsthm}
\usepackage{multirow}
\usepackage{footnote}
\usepackage{hyperref}
\usepackage[square,sort,comma,numbers]{natbib}
\usepackage{hypernat}
\usepackage{graphicx}
\usepackage[
per-mode=fraction,
separate-uncertainty,
retain-unity-mantissa=false,
product-units=power,
table-alignment=center,
detect-all,
binary-units=true,
exponent-product=\cdot,
exponent-base=10
]{siunitx}
\usepackage{color}

\usepackage{changes}
\newcommand{\stkout}[1]{\ifmmode\text{\sout{\ensuremath{#1}}}\else\sout{#1}\fi}
\setdeletedmarkup{\stkout{#1}}

\title{{Rapid, High-resolution and Distortion-free $R_{2}^{*}$ Mapping of Fetal Brain using Multi-echo Radial FLASH and Model-based Reconstruction\footnote{Part of this work has been presented at the ISMRM, 2024, Singapore and ISMRM 2025, Honolulu.}}}

\newcommand{\authorA}{Xiaoqing Wang}
\newcommand{\authorB}{Hongli Fan}
\newcommand{\authorC}{Zhengguo Tan}
\newcommand{\authorD}{Serge Vasylechko}
\newcommand{\authorE}{Edward Yang}
\newcommand{\authorF}{Ryne Didier}
\newcommand{\authorG}{Onur Afacan}
\newcommand{\authorH}{Martin Uecker}
\newcommand{\authorI}{Simon K. Warfield}
\newcommand{\authorJ}{Ali Gholipour}

\newcommand{\affilA}{Department of Radiology, Boston Children's Hospital, Harvard Medical School, Boston, Massachusetts, USA}
\newcommand{\affilB}{Siemens Medical Solutions, Boston, Massachusetts, USA}
\newcommand{\affilD}{Institute of Biomedical Imaging, Graz University of Technology, Graz, Austria}
\newcommand{\affilC}{Department of Radiology, University of Michigan, Ann Arbor, Michigan, USA}
\newcommand{\affilE}{Department of Radiological Sciences, University of California Irvine, Irvine, California, USA}
\newcommand{\affilF}{ Department of Electrical Engineering and Computer Science, University of California Irvine, Irvine, California, USA}

\newcommand{\corAdress}{Xiaoqing Wang, Department of Radiology, Boston Children's Hospital, Harvard Medical School, 300 Longwood Avenue, 02115, Boston, MA, USA. }

\newcommand{\corMail}{xiaoqing.wang@childrens.harvard.edu}

\iftoggle{ARXIV}{
	\author[1]{\authorA \thanks{\corAdress \corMail}}
}

\author[1]{\authorA}
\author[2]{\authorB}
\author[3]{\authorC}
\author[1]{\authorD}
\author[1]{\authorE}
\author[1]{\authorF}
\author[1]{\authorG}
\author[4]{\authorH}
\author[1]{\authorI}
\author[1,5,6]{\authorJ}

\affil[1]{\affilA}
\affil[2]{\affilB}
\affil[3]{\affilC}
\affil[4]{\affilD}
\affil[5]{\affilE}
\affil[6]{\affilF}

\begin{document}
	
%
	
%
%
%
%
%
%
%

	\maketitle
	
%
%
%
%
%
%
	
	\section*{Abstract}
	
	\noindent     
    \textbf{Purpose}: To develop a rapid, high-resolution and distortion-free technique for simultaneous water-fat separation, $R_{2}^{*}$ and $B_{0}$ mapping of the fetal brain at 3T. 

\noindent \textbf{Methods}: 
A 2D multi-echo radial FLASH sequence with blip gradients is adapted for data acquisition during maternal free breathing. A calibrationless model-based reconstruction with sparsity constraints is developed to jointly estimate water, fat, $R_{2}^{*}$ and $B_{0}$ field maps directly from k-space. This approach was validated and compared to reference methods using numerical and NIST phantoms and data from nine fetuses between 26 and 36 weeks of gestation age.
    
	
%
	\noindent
	\textbf{Results}: 
    Both numerical and experimental phantom studies confirm good accuracy and precision. In fetal studies, model-based reconstruction yields quantitative $R_{2}^{*}$ values in close agreement with those from a parallel imaging compressed sensing (PICS) technique using Graph Cut (intra-class correlation coefficient [ICC] = 0.9601), while providing enhanced image detail. Repeated scans confirm good reproducibility (ICC = 0.9213). Compared to multi-echo EPI, the proposed radial technique produces higher-resolution (1.1 $\times$ 1.1 $\times$ 3 mm$^{3}$ vs. 2-3 $\times$ 2-3 $\times$ 3 mm$^{3}$) $R_{2}^{*}$ maps with reduced distortion. Despite of differences in motion, resolution and distortion, $R_{2}^{*}$ values are comparable between the two acquisition strategies (ICC = 0.8049). Additionally, the proposed approach enables synthesis of high-resolution and distortion-free $R_{2}^{*}$-weighted images. 

	\noindent \textbf{Conclusion}: 
    This study demonstrates the feasibility of using multi-echo radial FLASH combined with calibrationless model-based reconstruction for motion-robust, distortion-free $R_{2}^{*}$ mapping of the fetal brain at 3T, achieving a nominal resolution of $1.1 \times 1.1 \times 3$ mm$^{3}$ within 2 seconds per slice.

	\noindent
	\textbf{Keywords}: $R_{2}^{*}$ mapping, fetal MRI, distortion-free, multi-echo radial FLASH,  model-based reconstruction
	
	\clearpage

	\section*{Introduction}
	\label{sec:introduction}
The importance of quantitative $R_{2}^{*}$ (where $R_{2}^{*} = 1/T_{2}^{*}$) mapping of the fetal brain has been increasingly recognized. For example, changes in $R_{2}^{*}$ values across gestational age provide a quantitative measure of early brain development \cite{vasylechko2015t2}. Furthermore, $R_{2}^{*}$ mapping and $R_{2}^{*}$-weighted imaging are valuable in identifying intracranial hemorrhage in the fetal brain \cite{sanapo2017fetal, epstein2021prenatal}. The quantitative values are also playing an important role for optimizing $R_{2}^{*}$-weighted functional fetal MRI \cite{rivkin2004prolonged, blazejewska20173d, zhao2024age}.  However, obtaining accurate and high-resolution $R_{2}^{*}$ mapping of the fetal brain is challenging due to motion caused by maternal respiration and unpredictable fetal movements \cite{gholipour2014fetal, calixto2024advances}. As a result, single-shot sequences, particularly single-shot 2D multi-echo Echo-Planar Imaging (EPI)-based approaches \cite{vasylechko2015t2, blazejewska20173d,vasylechkofetal3T, afacan2019fetal, turk2022change, nichols2024t2}, are typically used for $R_{2}^{*}$ quantification of fetal brain. These techniques were initially developed for 1.5 T \cite{vasylechko2015t2} and 3.0 T \cite{blazejewska20173d,vasylechkofetal3T}, with recent adaptations for 0.55 T \cite{zhang2024structural}. While relatively higher resolution and signal-to-noise ratio (SNR) imaging is achievable at higher fields, low field (e.g., 0.55 T) imaging has shown reduced distortion artifacts for the EPI readout \cite{aviles2023reliability}, which is attributed to reduced field inhomogeneity and smaller $R_{2}^{*}$s (i.e., longer $T_{2}^{*}$s). Consequently, quantitative $R_{2}^{*}$ mapping of fetal body organs has also been reported at 0.55 T \cite{payette2024fetal}.

Despite the scan efficiency of EPI, its prolonged readout makes EPI susceptible to geometric distortion caused by $B_{0}$ field inhomogeneity, particularly at higher field strengths. Additionally, in the multi-echo EPI sequence, the extended readout time necessitates a trade-off between imaging speed (i.e., short echo times and small inter-echo spacing) and spatial resolution due to $T_{2}^{*}$ decay \cite{wang2019echo}. For instance, the commonly reported spatial resolution for fetal brain imaging is $3 \times 3 \times 3$ mm$^{3}$ \cite{vasylechkofetal3T, blazejewska20173d, turk2022change}, which may limit its usefulness in clinical diagnosis where high-resolution imaging is required \cite{sanapo2017fetal, epstein2021prenatal, calixto2024advances, story2024functional}.

Radial acquisition is an alternative sampling strategy that has gained significant interest in the past decade due to its tolerance to data undersampling and robustness against motion \cite{Peters_Magn.Reson.Med._2000, Block_Magn.Reson.Med._2007, Feng_Magn.Reson.Med._2014, Block_J.KoreanSoc.Magn.Reson.Med._2014}. It has been applied to imaging children with reduced sedation \cite{kecskemeti2018robust, armstrong2018free,hu20193d} and in free-breathing fetal studies \cite{liu2014fast, sun2020feasibility,armstrong20193d}. Stack-of-stars multi-echo radial fast low-angle shot (FLASH) sequence has also been used for quantitative $R_{2}^{*}$ mapping in adult abdominal imaging \cite{Benkert_Magn.Reson.Med._2017, Schneider_Magn.Reson.Med._2020, liu2021robust, Tan_IEEE_TMI_2023,shih2023uncertainty,  zhong2024accelerated} and the fetal placenta \cite{armstrong20193d}. However, the unpredictable motion of the fetal brain, combined with maternal motion and motion-induced phase errors, poses significant challenges for applying 3D sequences in general (i.e., including both Cartesian or non-Cartesian sequences) to quantitative imaging of the fetal brain.

Alongside motion-robust sequence design, advanced image reconstruction is essential for efficient quantitative imaging. Reconstruction techniques that incorporate prior signal model information to constrain parameter space have been developed \cite{petzschner_Magn.Reson.Med._2011, huang_Magn.Reson.Med.2012, Zhao_Magn.Reson.Med._2015, Tamir_Magn.Reson.Med._2017, Block_IEEETrans.Med.Imaging_2009, Fessler_IEEESignalProcess.Mag._2010, Wang_Philos.Trans.R.Soc.A._2021}. Among these, nonlinear model-based reconstruction techniques  \cite{Wang_Philos.Trans.R.Soc.A._2021, scholand2023quantitative} are highly efficient. These techniques incorporate complex spin dynamics directly in the reconstruction.
By formulating reconstruction as a nonlinear inverse problem, model-based reconstruction can estimate physical quantitative maps from undersampled k-space data without intermediate reconstruction or pixelwise fitting. Advanced regularization techniques, such as sparsity constraints \cite{Lustig_Magn.Reson.Med._2007}, further enhance precision in quantitative mapping. Recently, this approach has been extended to reconstruct water, fat, and $R_{2}^{*}$ maps from undersampled 3D multi-echo FLASH for liver imaging \cite{Benkert_Magn.Reson.Med._2017, Schneider_Magn.Reson.Med._2020}, also enabling additional $B_{0}$ estimation \cite{Tan_IEEE_TMI_2023}.

Building on the ideas above, this work aims to develop a rapid quantitative $R_{2}^{*}$ mapping of fetal brain utilizing a 2D multi-echo radial FLASH sequence and a calibrationless model-based reconstruction. While the radial sequence provides motion robustness and efficient k-space coverage for fetal imaging, the model-based reconstruction estimates quantitative maps directly from undersampled k-space. This combination enables high-resolution and distortion-free quantitative $R_{2}^{*}$ mapping ($1.1 \times 1.1 \times 3$ mm$^{3}$) of fetal brain in \textbf{two seconds per slice}. Validations and comparison to reference methods have been performed on numerical simulations, experimental phantom, and nine fetuses each scanned at an age between 26 to 36 weeks of gestation.

	\section*{Methods}
	\label{sec:theory}
	\subsection*{Sequence Design}
	\label{subsec:sequence}
	A 2D multi-echo radial FLASH sequence is adapted for data acquisition. Similar to \cite{Tan_Magn.Reson.Med._2019}, radial spokes are designed to rotate along the echo dimension
	using blip gradients, enabling an efficient
	k-space coverage (Supporting Information Figure S1).
 The distribution of spokes is designed in a way that radial lines from several excitations (e.g., 3) and all echoes are equally distributed \cite{Tan_Magn.Reson.Med._2019} in one k-space, with an angle $\theta_{l,m} = 2 \pi /
	 (N_\text{E} \cdot N_\text{S}) \cdot [(l-1) \cdot N_\text{E} + m-1]$ for the $l$th TR and the $m$th echo. 
	 $N_\text{E}$ and $N_\text{S}$ are the number of echoes and shots (TRs)
	 per k-space, respectively. Spokes acquired in consecutive
	 k-space frame are then rotated by a small golden-angle ($\approx 68.75^{\circ}$) with respect to
	 the previous one \cite{Wundrak_Magn.Reson.Med._2016_2} to enable a complementary coverage of k-space. Since the $R_{2}^{*}$ values of fetal brain are reported to be much smaller than those of adult brains \cite{rivkin2004prolonged}, the number of echoes is extended from 7 \cite{wang2024model} to 35 to enable a robust $R_{2}^{*}$ estimation, while reducing the risk of phase wrapping at later echoes \cite{Tan_IEEE_TMI_2023, zhang2019neonate}. The above choice also aligns well with a recent neonatal brain study \cite{zhang2019neonate}. 
\begin{figure}[H]
	\centering
	\includegraphics[width=0.9\textwidth]{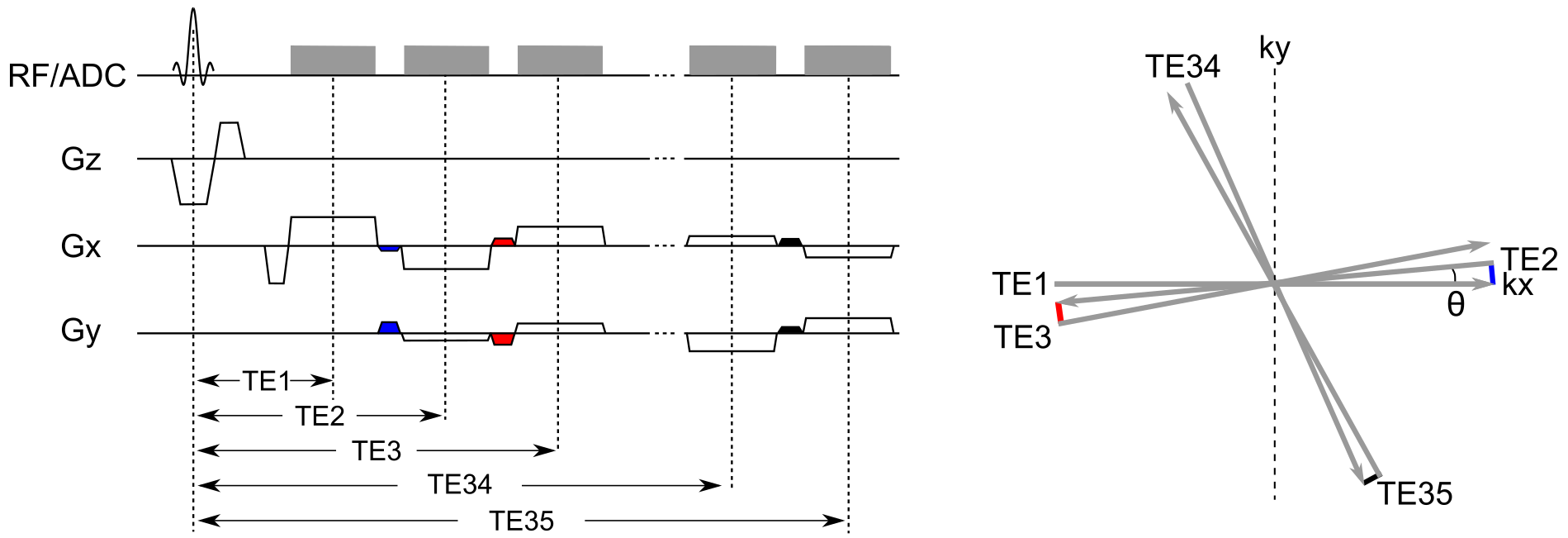}
	\caption*{Supporting Information Figure S1. Schematic diagram of the 2D multi-echo radial FLASH sequence (the first three echoes and the last two echoes are shown). Blip gradients (blue, red and black regions) are introduced among echoes to enable a complementary k-space coverage. $\theta$ is determined in a way that spokes from all echoes and 3 TRs are equally distributed. I.e., with 35 echoes and 3 TRs,  $\theta = 360^{\circ} / (35\times3)$.} 
\end{figure}

\subsection*{Signal Equation and Model-based Reconstruction}
Although the fetal brain contains minimal fat, surrounding tissues, such as maternal body tissue, include fat. To account for this, we construct the signal equation as follows \cite{Tan_IEEE_TMI_2023}:
	\begin{equation}
	\label{eq::sigmod}
M_{\text{TE}_{m}}=
({W} +{F}\cdot z_{m})
\cdot \exp\big(\text{TE}_{m}\cdot i2\pi \cdot {f_{B_{0}}}\big) \cdot \exp\big(- \text{TE}_{m} \cdot {R_{2}^{*}}\big)
\end{equation}
with $W$ and $F$ being the water and fat components, respectively; $z_{m}$ is the summarized 6-peak fat spectrum \cite{Yu_Magn.Reson.Med._2008} at echo time $\text{TE}_{m}$; and $f_{B_{0}}$ and $R_{2}^{*}$ are the corresponding field map and relaxation rate, respectively.
The estimation of the unknowns $(W, F, R_{2}^{*}, f_{B_{0}})^{T}$ is then formulated as a nonlinear inverse problem; i.e., by combining the above physical model with the parallel imaging equation 
\cite{Pruessmann_Magn.Reson.Med._1999, Uecker_Magn.Reson.Med._2008}, 
we construct a nonlinear forward operator $F$, which maps the unknowns 
in Equation (\ref{eq::sigmod}) and the unknown coil sensitivities $C$ to the 
acquired multi-channel data $y$ at $\text{TE}_{m}$, i.e., 
\begin{equation}
	F: x \mapsto y = \mathcal{P} \mathcal{F}C \cdot M_{ \text{TE}_{m}}(x_{p})~.
	\label{eq::forward}
\end{equation}
Here, $\mathcal{P}$ is the sampling pattern and $\mathcal{F}$ is 
the Fourier transform.  By defining $x_{c} = (c_{1}, \dots, c_{k}, \dots,
c_{K})^{T}$, with $c_{k}$ the individual $k$th coil sensitivity map, 
the vector of unknowns in Equation (\ref{eq::forward}) is $x = (x_{p}, x_{c})^{T}$. 
The estimation of $x$ is then formulated as an optimization problem, i.e.,
\begin{equation}
\hat{x} = \text{argmin}_{ x \in D} \frac{1}{2}\sum_{\text{TE}}\big\|P\mathcal{F}C\cdot M_{ \text{TE}_{m}}(x)- Y_{\text{TE}_{m}}\big\|_{2}^{2} + \alpha R(x).\
\end{equation}
Here, D is a convex set, ensuring non-negativity of $R_{2}^{*}$. $R(\cdot)$ is the regularization term for both parameter 
maps and coil sensitivity maps with $\alpha$ the regularization parameter. In particular, we use joint 
$\ell_{1}$-Wavelet sparsity constraint \cite{Wang_Magn.Reson.Med._2018}
on $(W, F, R_{2}^{*})^{T}$ 
to exploit sparsity and correlations between maps
and Sobolev regularization on 
the $f_{B_{0}}$ map \cite{Tan_IEEE_TMI_2023, wang2024model} and the coil sensitivity maps \cite{Uecker_Magn.Reson.Med._2008} to enforce smoothness. The Sobolev regularization reads:
\begin{equation}
R(\cdot) = \|(1+s\|\vec{k}\|^{2})^{l/2}\mathcal{F}\{\cdot\}\|^{2}
\end{equation}
where $\|\vec{k}\|$ defines the distance to the k-space center, $s$ and $l$ are constants. The above optimization problem is solved by IRGNM-FISTA 
\cite{Wang_Magn.Reson.Med._2018} using the Berkeley Advanced Reconstruction Toolbox (BART) \cite{Uecker__2015}.

\subsection*{Numerical Simulations}
To validate the accuracy of the proposed approach, a numerical phantom with ten circular tubes and a background was simulated. The $R_{2}^{*}$ values were set to be from 10 s$^{-1}$ to 200 s$^{-1}$ (i.e., $T_{2}^{*}$ from 10 ms to 200 ms with a step size of 20 ms). The off-resonance ranged from -50 Hz to 50 Hz with a step size of 10 Hz. The fat fraction was set to be from 5$\%$ to 95$\%$ with a step size of 10$\%$. The k-space data was derived from the analytical Fourier representation of an ellipse assuming an array of eight circular receiver coils surrounding the phantom. The 2D multi-echo radial FLASH sequence described in the Sequence Design section was used to sample the simulated k-space with a base resolution of 192 pixels covering a field of view (FOV) of 128 mm. The other sequence parameters are the same as those listed in the following Experiments section.
Complex white Gaussian noise (standard deviation = 0.1) was added to the simulated k-space data to mimic noise levels typical of modern 3T MRI scanners. Moreover, simulations with different degrees of noise were performed
for appraising the achievable reconstruction accuracy against noise. 

\subsection*{Experiments}
All MRI experiments were conducted on a Magnetom Prisma 3T scanner (Siemens Healthineers, Erlangen, Germany) during maternal free breathing. The study was approved by the Institutional Review Board, and written informed consent was obtained from all participants.
Validation was first performed using the $T_{1}$ spheres of a NIST phantom \cite{stupic2021standard}. Phantom scans utilized a 64-channel head/neck coil to achieve high SNR, while fetal imaging employed the standard 30-channel abdominal coil provided by the vendor, commonly used in our research scanner for fetal studies. 
Nine pregnant female subjects (35 $\pm$ 4 years old; fetuses: 31.6 $\pm$ 3.6 weeks)
without known illness were enrolled and scanned. Standard Half Fourier Single-shot Turbo spin-Echo (HASTE) images were acquired first for each subject in three (axial, coronal, and sagittal) orientations of the fetal brain with a FOV of  256 × 256 mm$^{2}$, matrix size= 256 $\times$ 256, slice thickness = 2 mm, and a total acquisition time of 1-1.5 second per slice. Radial fetal scans were performed in the axial orientation with the following acquisition parameters: FOV = 256 $\times$ 256 mm$^{2}$, matrix
size= 224 $\times$ 224, slice thickness = 3 mm, 35 echoes with TR = 68.3 ms,
TE$_{1}$/$\delta$TE/TE$_{35}$ = 2.37/1.88/66.90 ms, flip angle (FA) = $20^{\circ}$,
bandwidth = 740 Hz/pixel, and 30 RF excitations with 1050 radial acquired
spokes for all echoes.  For quantitative comparison, multi-echo EPI images were acquired with FOV = 256 × 256 mm$^{2}$, matrix size= 96-128 × 96-128, slice thickness = 3 mm,
TEs = (23.4-29.8, 74.90-77.48, 126.38-147.20, 177.88-207.46) ms. Quantitative EPI $R_{2}^{*}$ maps were generated via pixelwise magnitude fitting to the exponential model $y_{m} = \rho \cdot \exp(-\text{TE}_{m} \cdot R_{2}^{*})$. Both EPI and radial scans were acquired with 16-20 slices in an interleaved manner to cover a majority of the fetal brain. Radial scans were able to be repeated in seven of the nine subjects to assess the repeatability of the proposed method.

Additionally, for the phantom study, a vendor-provided 3D Cartesian multi-echo sequence was used for reference with these parameters:
FOV = 256 $\times$ 256 mm$^{2}$, matrix size= 224 $\times$ 224, slice thickness = 3 mm with 30 slices, 11 echoes with TR = 65 ms,
TE$_{1}$/$\delta$TE/TE$_{11}$ = 6.0/5.5/61.0 ms, FA = $15^{\circ}$,
bandwidth = 300 Hz/pixel, and acceleration factor 2. The total acquisition time was 4:17 min. The Cartesian $R_{2}^{*}$ and $B_{0}$ maps were estimated by fitting the multi-echo complex images to the modified signal model (Equation \ref{eq::sigmod}), excluding the fat component. Noteworthy, for fetal imaging, the flip angle was set near the Ernst angle to maximize the SNR in the multi-echo FLASH acquisition \cite{bernstein2004handbook}. This was based on typical fetal brain $T_1$ values and the sequence parameters used in this study, and is consistent with values reported in recent neonatal imaging work \cite{jang2025quantitative}. For the NIST phantom, which contains a broad range of $T_1$ values, the optimal flip angle varies by tube. In accordance with recommendations from a recent consensus paper \cite{qsm2024recommended}, a flip angle of $15^\circ$ was used for the 3D Cartesian sequence in phantom experiments.

\subsection*{Iterative Reconstruction}
All iterative reconstructions were performed offline using BART \cite{Uecker__2015}. The multi-echo radial FLASH datasets from multiple receiver coils were first corrected for gradient delay errors using RING \cite{Rosenzweig_Magn.Reson.Med._2018a} and then compressed to 12 virtual coils via principal component analysis. The data and sampling trajectory were subsequently gridded onto a Cartesian grid, where all iterative steps were carried out using FFT-based convolutions with the point-spread function \cite{Wajer__2001, Uecker_NMRBiomed._2010, wang2023free_MRM}. The regularization parameter $\alpha$ was initialized at 1.0 and reduced by a factor of three with each Gauss-Newton iteration, following $\alpha_{n+1} = \max(\alpha_{\text{min}}, (1/3)^n\cdot \alpha_{0})$ with $\alpha_{0}=1.0$. The regularization value $\alpha_{\text{min}}$ was then chosen based on visual inspection to balance noise suppression and quantitative accuracy. Similar to our previous study \cite{Tan_IEEE_TMI_2023}, the constants $s$ and $l$ for $f_{B_{0}}$ regularization were set to 22 mm$^{2}$ and 4, respectively, to balance field map smoothness and accuracy.
The model-based iterative reconstruction was executed on a GPU with 48 GB of memory (RTX A6000, NVIDIA, Santa Clara, CA), with a computation time of 5–10 minutes per dataset. For comparison, the same multi-echo datasets were jointly reconstructed using the parallel imaging and compressed sensing (PICS) method, with coil sensitivity maps estimated from the first echo and joint sparsity constraints applied across spatial and echo dimensions. After image reconstruction, quantitative water, fat, $R_{2}^{*}$, and 
$B_{0}$ maps were estimated using the Graph Cut technique \cite{hernando2010robust}, available in the ISMRM water-fat toolbox \cite{Hu_Magn.Reson.Med._2012}.

\subsection*{Quantitative Analysis}

All quantitative results are reported as mean $\pm$ standard deviation (SD). 
 For phantom studies, regions of interest (ROIs) were carefully placed at the center of each tube to minimize potential partial volume effects, using the arrayShow tool \cite{Sumpf__2013} implemented in MATLAB (MathWorks, Natick, MA). For in vivo studies, ROIs were manually drawn into the frontal white matter (FWM), thalamic gray matter (THA), and occipital white matter (OWM) regions of the central-slice fetal brain $R_{2}^{*}$ maps \cite{vasylechko2015t2} utilizing the same arrayShow tool. Bland–Altman analyses were used to compare 
ROI-based mean quantitative $R_{2}^{*}$
values between the proposed technique and reference methods, i.e., PICS with Graph Cut for the same radial data and the multi-echo EPI approach. The Intra-Class Correlation Coefficient (ICC) was further utilized to assess both the agreement between the proposed technique and reference methods, and the reproducibility between repeated scans of the proposed radial approach.

\section*{Results}
\label{sec:results}



\subsection*{Phantom Validation}

We first validated the proposed technique on a numerical phantom, which offers a broad range of ground-truth quantitative values under noisy conditions.
Figure 1 (top) shows water, fat fraction (FF), $R_{2}^{*}$, and $B_{0}$ maps obtained from the model-based reconstruction with a 2-second multi-echo radial FLASH acquisition. Figure 1 (bottom) compares ROI-analyzed quantitative values with the ground truth. 
The mean differences are $-0.03 \pm 0.05$ $\%$, $-0.17 \pm 0.08$ $s^{-1}$ and $0.01 \pm 0.07$ Hz for FF, $R_{2}^{*}$, and $B_{0}$, respectively. The low mean differences indicate good quantitative accuracy of the proposed method. 
The Supporting Information Figure S2 (top) illustrates model-based reconstructed $R_{2}^{*}$ maps under varying noise levels—low, medium, and high. Despite the increased noise, the $R_{2}^{*}$ maps remain visually comparable. This observation is quantitatively confirmed by the Bland–Altman analysis in Supporting Information Figure S2 (bottom) and the data in Supporting Information Table S1, which show good agreement in mean values with the ground truth, despite the expected increase in standard deviations with higher noise levels. These findings demonstrate the robustness of the proposed method against noise. 
\begin{figure}[H]
	\centering
	\includegraphics[width=0.9\textwidth]{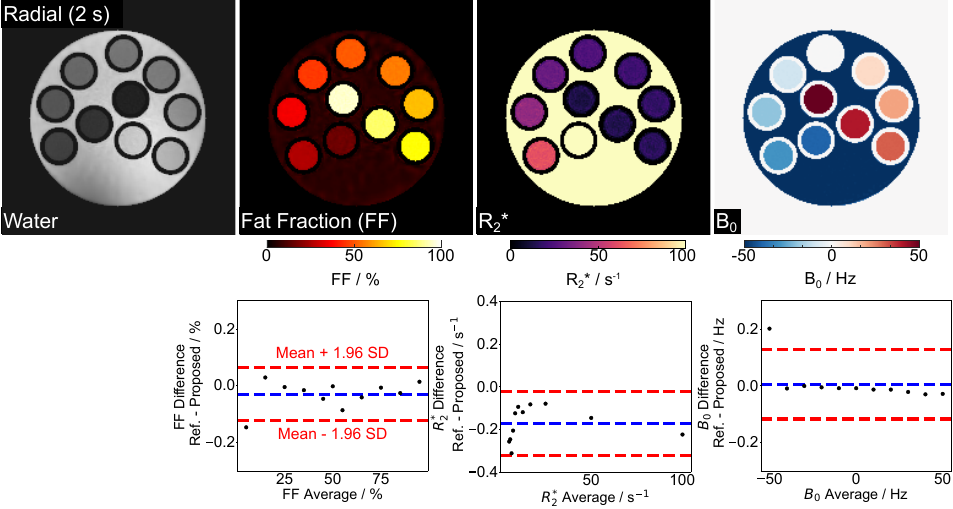}
	\caption*{ Figure 1. (Top) Model-based estimated water, fat fraction, $R_{2}^{*}$, and $B_{0}$ field maps using a 2-second multi-echo radial FLASH sequence for a numerical phantom. (Bottom) Bland-Altman plots comparing the ROI-analyzed mean quantitative values to the ground truth. The mean differences are $-0.03 \pm 0.05$ $\%$, $-0.17 \pm 0.08$ $s^{-1}$ and $0.01 \pm 0.07$ Hz for FF, $R_{2}^{*}$ and $B_{0}$, respectively.} 
\end{figure}

Figure 2 presents NIST $R_{2}^{*}$ (top) and $B_{0}$ (bottom) maps generated by the proposed method and a 3D Cartesian reference. Note that here a 3-parameter model (i.e., excluding fat in Equation \ref{eq::sigmod}) was employed in the reconstruction as there is no known fat component in the NIST phantom. Despite phase wrap differences around the central top two tubes on the $B_{0}$ maps, both visual inspection and quantitative ROI analysis demonstrate good agreement: The mean $R_{2}^{*}$ difference is $0.6 \pm 2.4$ s$^{-1}$ for $R_{2}^{*}$ ranging from 4 s$^{-1}$ to 60 s$^{-1}$. 

\begin{figure}[H]
	\centering
\includegraphics[width=0.8\textwidth]{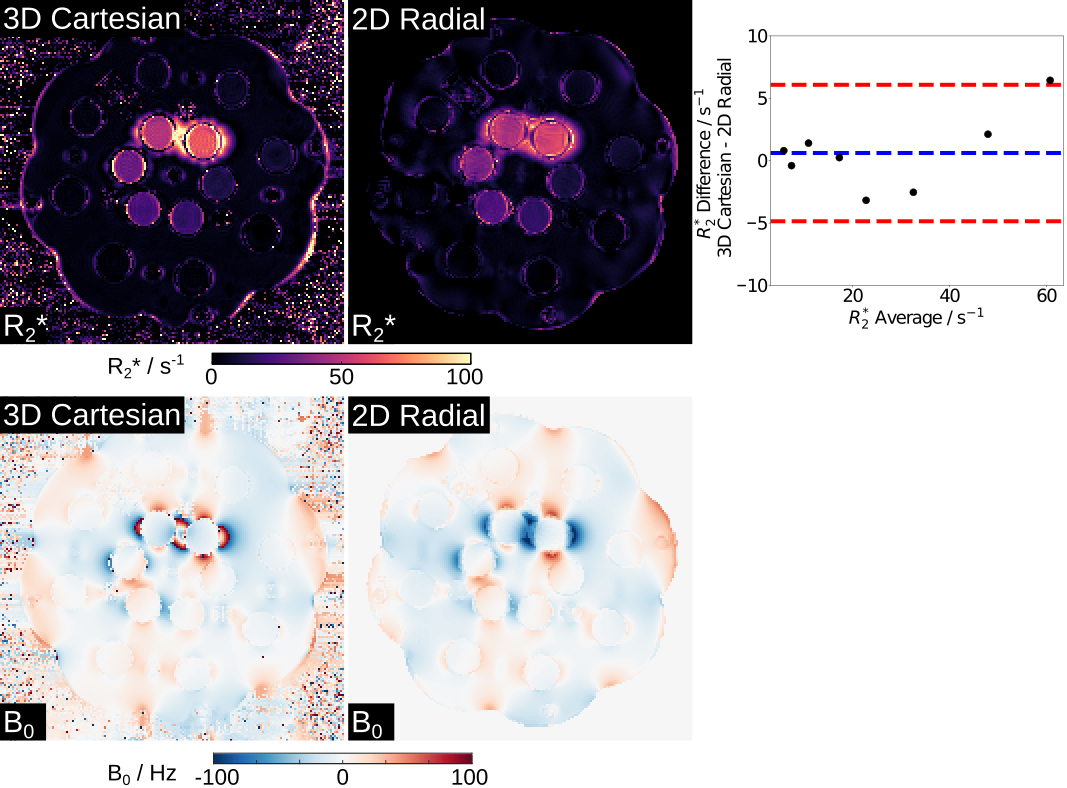}
\caption*{Figure 2. Model-based estimated (top) $R_{2}^{*}$ and (bottom) $B_{0}$ maps and their comparison to the 3D Cartesian references of the NIST phantom (T1 sphere).
	(Top right) Bland–Altman plots comparing the ROI-analyzed mean quantitative $R_{2}^{*}$ values to the references. The mean difference is $0.6 \pm 2.8$ s$^{-1}$. Note
	that the 3D Cartesian reference acquisition time is 4:17 min, while the 2D radial sequence requires 2 seconds per slice.}
\end{figure}


\subsection*{Fetal Studies}


Figure 3 illustrates the impact of the regularization parameter $\alpha_{\text{min}}$ on fetal brain $R_{2}^{*}$ maps from two representative subjects. Supporting Information Figure S3 presents the corresponding quantitative values across six ROIs of each subject. As expected, lower $\alpha_{\text{min}}$s lead to increased noise (higher standard deviation), while higher ones cause image blurring. An optimal value of $\alpha_{\text{min}} = 0.002$ was selected to balance noise suppression and preservation of anatomical detail. 

\begin{figure}[H]
	\centering
	\includegraphics[width=0.8\textwidth]{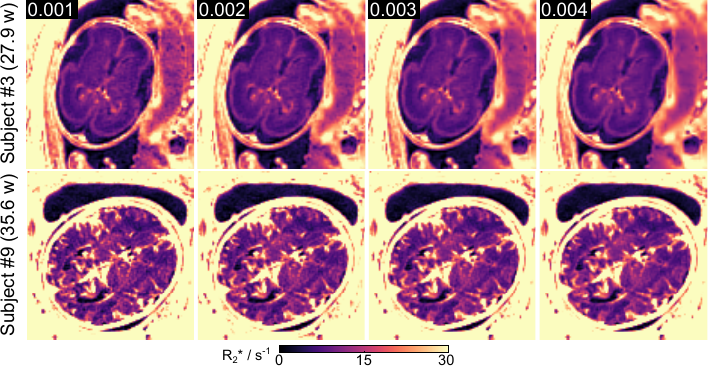}
	\caption*{Figure 3. Model-based fetal brain $R_{2}^{*}$ maps as a function of the regularization parameter $\alpha_{\text{min}}$ for two representative subjects. A value of 0.002 is utilized for all in vivo studies.} 
\end{figure}

With the above settings, Figure 4 (A) shows reconstructed water, fat, $R_{2}^{*}$, and $B_{0}$ maps obtained using the proposed model-based method and a PICS reconstruction with the Graph Cut technique on the same radial dataset. Visual inspection indicates good correspondence between the two methods.
Figure 4 (B) includes enlarged $R_{2}^{*}$  maps, synthesized $R_{2}^{*}$-weighted images at TE = 60 ms, and Bland-Altman plots comparing mean $R_{2}^{*}$  values for selected ROIs (white circles). 
Despite the proposed model-based method showing a better balance between preserving fine details and reducing noise in both $R_{2}^{*}$ maps and synthesized $R_{2}^{*}$-weighted images (black arrows), the low mean difference ($0.07 \pm 0.17$ s$^{-1}$) confirms strong quantitative agreement.
\begin{figure}[H]
	\centering
	\includegraphics[width=0.9\textwidth]{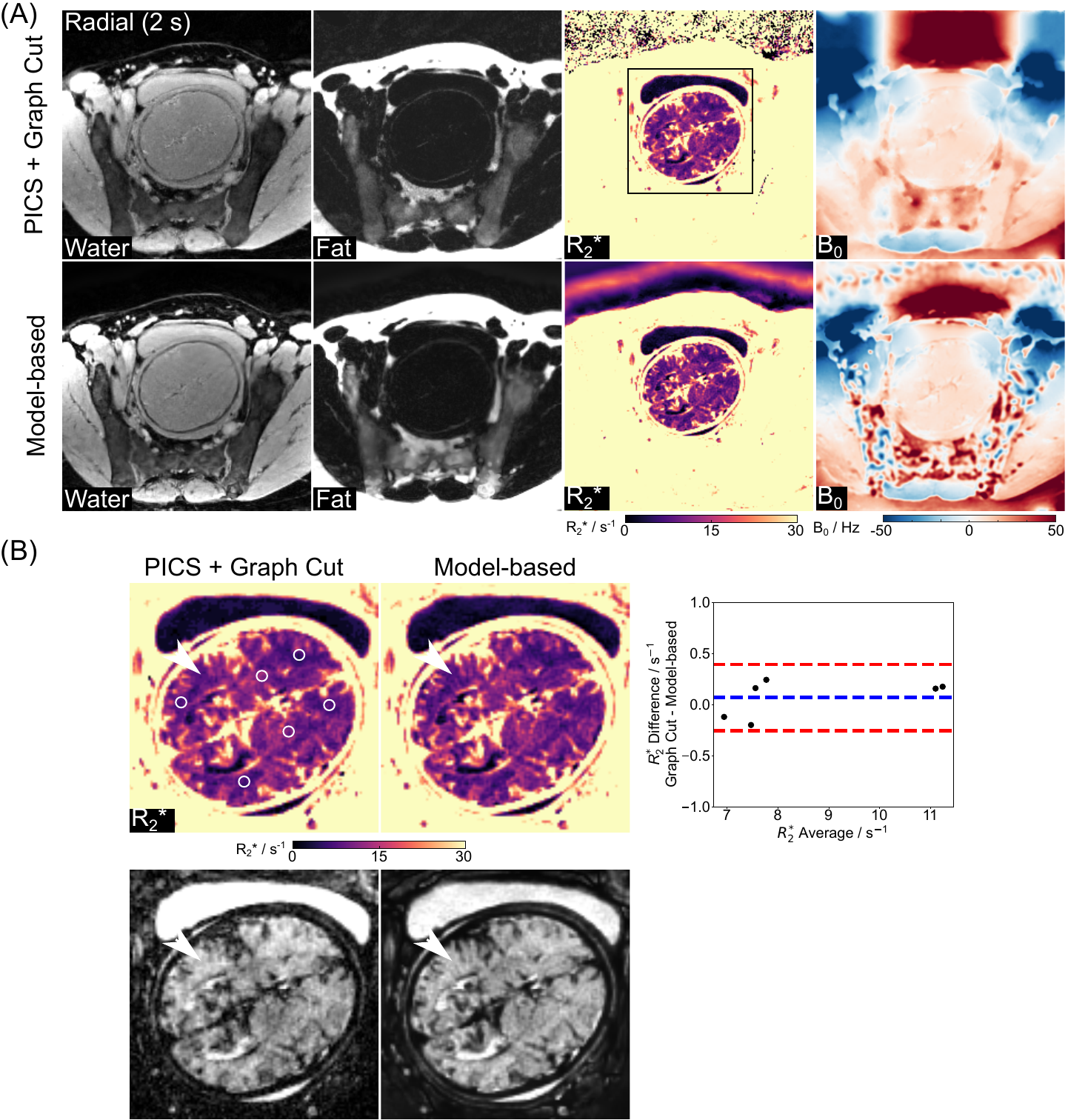}
	\caption*{Figure 4. (A). Model-based reconstructed water, fat, $R_{2}^{*}$, and $B_{0}$ maps and their comparison to a reference method (Parallel imaging compressed sensing
		with Graph Cut) utilizing the same radial data. (B).  Enlarged $R_{2}^{*}$ maps and $R_{2}^{*}$-weighted images (TE = 60 ms) and the Bland–Altman plots comparing the ROI-analyzed (white circles) mean quantitative $R_{2}^{*}$ values. The mean difference is $0.07 \pm 0.17$ s$^{-1}$ for all ROIs. White arrows indicate a better balance between preserving fine details and reducing noise of the model-based approach.}
\end{figure}

The above findings are further supported by comparisons across additional subjects and quantitative results shown in Figure 5 and the Supporting Information Figure S4. Figure 5 (A) highlights comparable image quality with enhanced details in the model-based reconstruction (white and black arrows), while Figure 5 (B) demonstrates small quantitative differences ($0.06 \pm 0.42$ $s^{-1}$, $0.11 \pm 0.51$ $s^{-1}$, and $-0.13 \pm 0.64$ $s^{-1}$ for FWM, THA and OWM, respectively) between the two reconstruction approaches for all nine subjects. The similar mean $R_{2}^{*}$ values shown in Table 1 and a high ICC of 0.9601 further indicate a strong agreement between the two methods. 
\begin{figure}[H]
	\centering
	\includegraphics[width=0.95\textwidth]{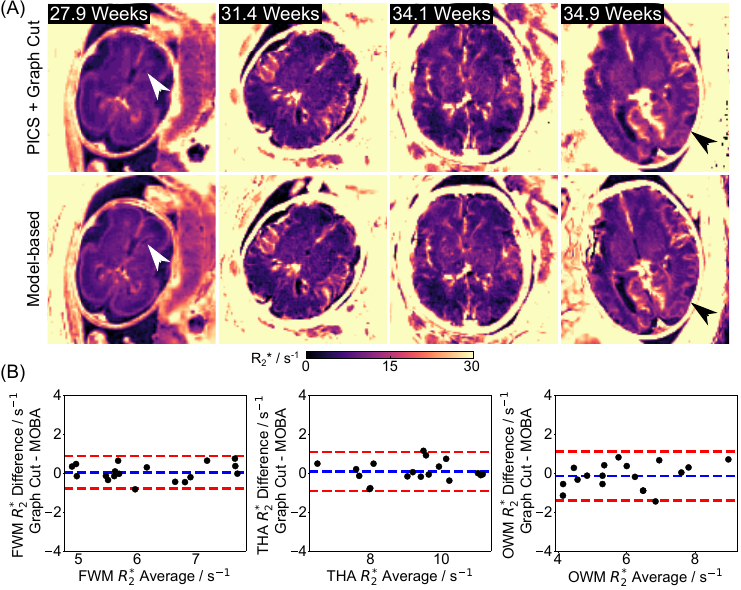}
	\caption*{Figure 5. (A). Comparison of quantitative fetal brain $R_{2}^{*}$ maps estimated using (top) PICS with Graph Cut and (bottom) model-based reconstruction in four representative subjects at different gestational ages. White and black arrows indicate improved image details by model-based reconstruction. (B). Bland–Altman plots comparing the mean quantitative $R_{2}^{*}$ values for all nine subjects in this study. The mean $R_{2}^{*}$ differences for FWM, THA and OWM are $0.06 \pm 0.42$ s$^{-1}$, $0.11 \pm 0.51$ s$^{-1}$, and $-0.13 \pm 0.64$ s$^{-1}$, retrospectively. } 
\end{figure}
Figure 6 (A) shows two repetitive fetal brain $R_{2}^{*}$ maps for seven (out of nine) subjects. Despite varying motion conditions that may have been different during the two scans, the quantitative maps are visually comparable. This observation is confirmed by the minimal quantitative differences observed in the selected ROIs (ICC: 0.9213), as shown in Figure 6 (B). 
\begin{figure}[H]
	\centering
	\includegraphics[width=0.9\textwidth]{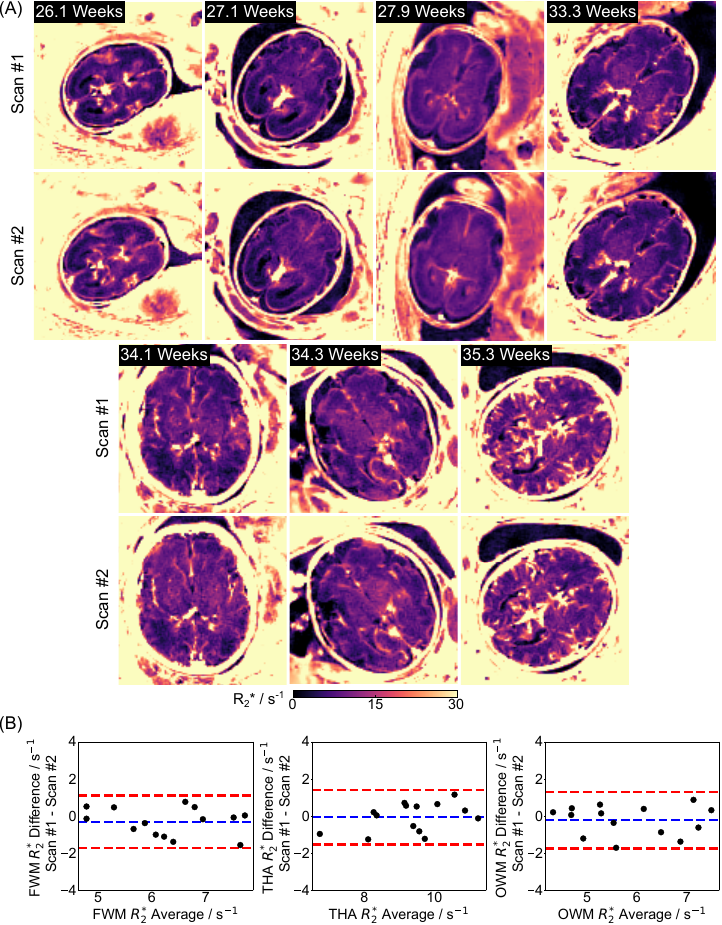}
	\caption*{Figure 6. (A). Quantitative fetal brain $R_{2}^{*}$ maps for two repeated scans across seven  (out of nine) subjects. (B). The scan-rescan $R_{2}^{*}$ differences are $-0.30 \pm 0.72$ s$^{-1}$, $-0.06 \pm 0.75$ s$^{-1}$ and $-0.22 \pm 0.78$ s$^{-1}$ for FWM, THA and OWM, respectively. Please note that repeat scans was not able to be performed on the other two subjects due to limited scan time during development.} 
\end{figure}

Figure 7 presents estimated radial water, $R_{2}^{*}$, and $B_{0}$ maps with model-based reconstruction, along with EPI $M_{0}$ and $R_{2}^{*}$ maps and T2-weighted HASTE images for two representative subjects (27.9 weeks and 35.6 weeks). Apart from motion-related differences, qualitative assessment demonstrates improved spatial resolution and reduced distortion (white arrows) by the proposed radial technique.
\begin{figure}[H]
	\centering
	\includegraphics[width=0.85\textwidth]{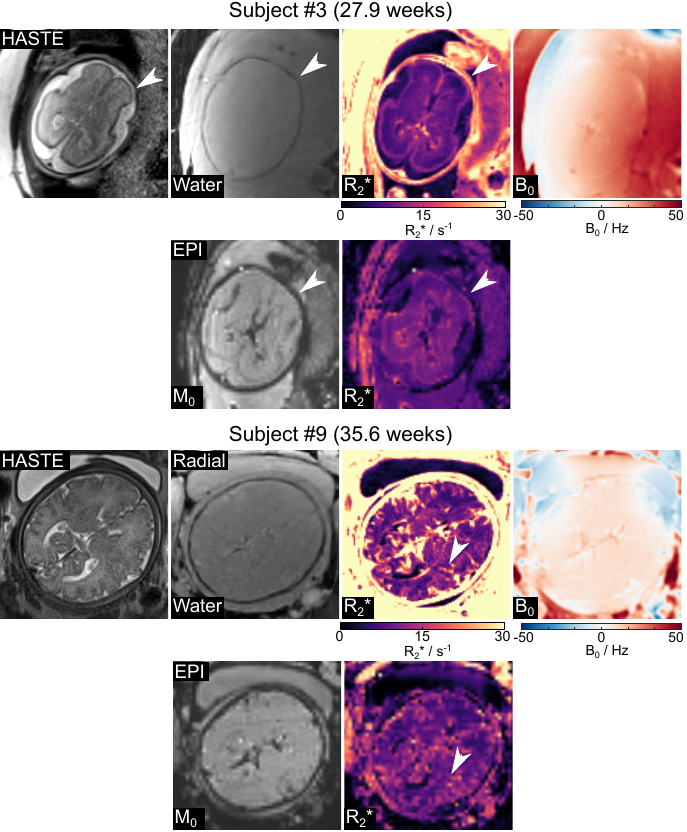}
	\caption*{Figure 7. HASTE images, model-based reconstructed water, $R_{2}^{*}$, and $B_{0}$ maps, along with a comparison to the EPI results for two representative subjects (27.9 weeks, top; 35.6 weeks, bottom). The radial acquisition demonstrates notably improved spatial resolution (both cases) and reduced distortions (Subject 3) compared to the EPI counterpart (white arrows). } 
\end{figure}

Figure 8 (A) compares quantitative $R_{2}^{*}$  maps generated from the radial and EPI techniques for the other seven subjects. Consistent with Figure 7, the radial $R_{2}^{*}$ maps exhibit enhanced delineation of small structures and less distortion (white and black arrows) compared to EPI maps. Figure 8 (B) shows ROI-analyzed quantitative values for both methods across all nine subjects. The mean differences for FWM, THA and OWM are $-0.54 \pm 1.00$ s$^{-1}$, $-0.74 \pm 1.10$ s$^{-1}$, and $0.34 \pm 0.90$ s$^{-1}$ between radial and EPI approaches. Additionally, Table 1 presents the mean $R_{2}^{*}$ values for all subjects. The EPI mean $R_{2}^{*}$ values are $5.7 \pm 1.1$ s$^{-1}$, $8.7 \pm 1.1$ s$^{-1}$, and $6.4 \pm 1.6$ s$^{-1}$, while the radial ones are $6.1 \pm 1.0$ s$^{-1}$, $9.1 \pm 1.3$ s$^{-1}$, $6.0 \pm 1.3$ s$^{-1}$ for FWM, THA and OWM, respectively. With an ICC of 0.8049, the above quantitative results suggests that the two methods yield comparable $R_{2}^{*}$ values. In addition to quantitative maps, Figure 9 demonstrates synthesized $R_{2}^{*}$-weighted images at TE= 70 ms (a typical value chosen for fetal functional MRI study) of the proposed radial approach and EPI methods as well as T2-weighted HASTE images across all subjects. In line with $R_{2}^{*}$ images, the contrast-weighted radial images show improved spatial resolution and reduced distortion compared to EPI (white arrows).  Moreover, radial FLASH images are less affected by B1 inhomogeneity than T2-weighted HASTE images, as they do not rely on a 180$^{\circ}$ refocusing pulse \cite{bernstein2004handbook}, offering an added value for high-resolution fetal imaging. 
 \begin{figure}[H]
 	\centering
 	\includegraphics[width=0.9\textwidth]{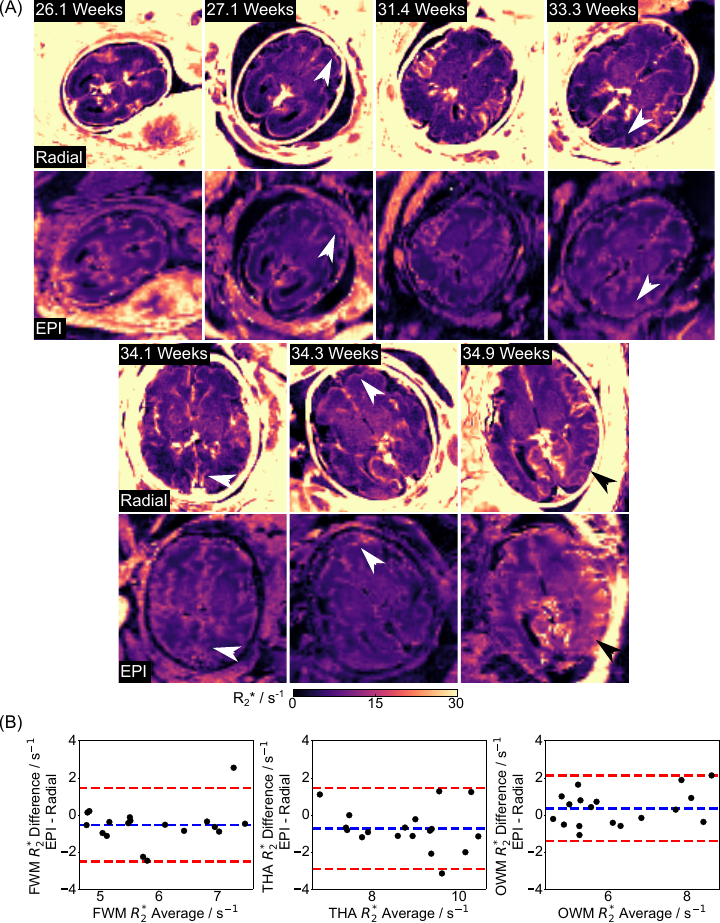}
 	\caption*{Figure 8. (A). Quantitative $R_{2}^{*}$ maps estimated with (top) multi-echo radial FLASH using model-based reconstruction and (bottom) EPI for the remaining seven subjects. Both white and black arrows indicate the enhanced delineation of small structures achieved by the radial approach.
 		(B). Bland–Altman plots comparing ROI mean quantitative $R_{2}^{*}$ values between the proposed technique and the EPI method for all nine subjects, showing a mean difference of $-0.54 \pm 1.00$ s$^{-1}$, $-0.74 \pm 1.10$ s$^{-1}$ and $0.34 \pm 0.90$ s$^{-1}$ for FWM, THA and OWM, respectively.} 
 \end{figure}

\begin{figure}[H]
	\centering
	\includegraphics[width=0.9\textwidth]{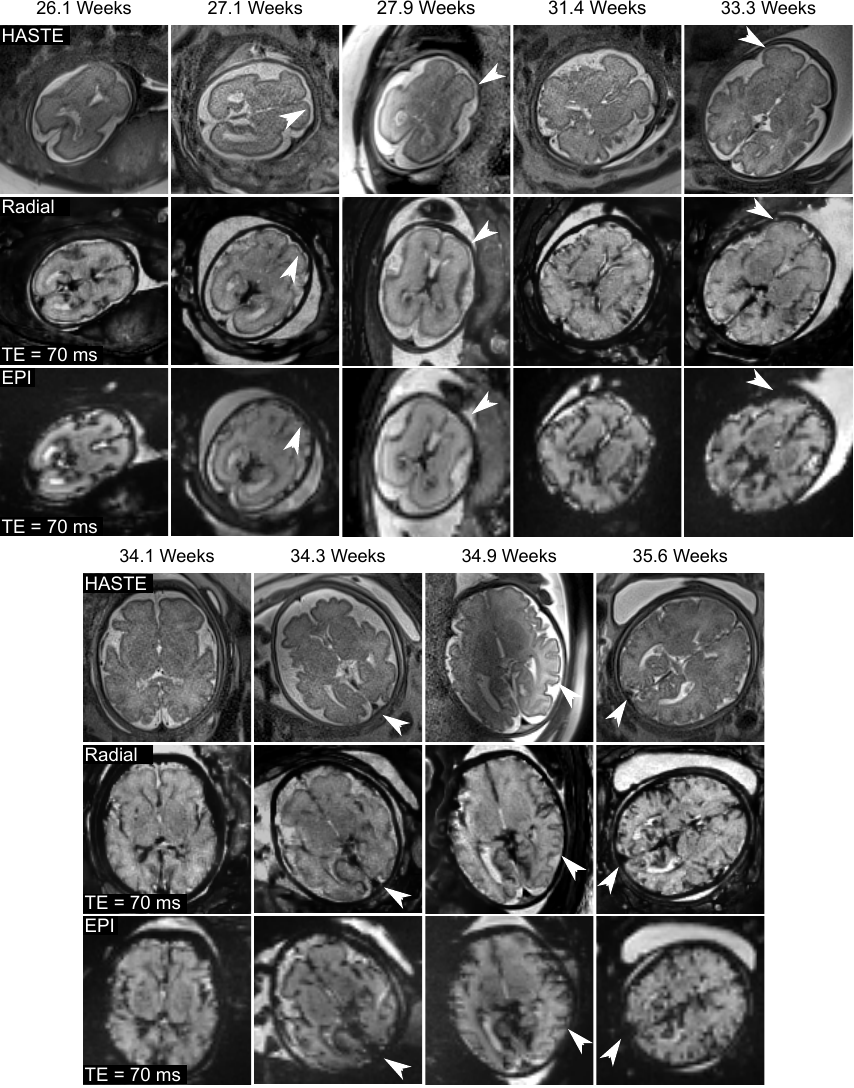}
	\caption*{Figure 9. (Top) HASTE images, synthesized $R_{2}^{*}$-weighted images at TE = 70 ms for (middle) radial and (bottom) EPI acquisitions for all nine subjects. The nominal spatial resolution for HASTE, radial FLASH and EPI are 1.0 $\times$ 1.0 $\times$ 2 mm$^{3}$, 1.1 $\times$ 1.1 $\times$ 3 mm$^{3}$, and 2-3 $\times$ 2-3 $\times$ 3 mm$^{3}$, respectively. The proposed radial approach achieves notably improved spatial resolution and reduced distortion compared to the EPI method, as indicated by the white arrows.} 
\end{figure}


The supporting information videos S1 and S2 provide the estimated radial $R_{2}^{*}$ maps and corresponding synthesized 
$R_{2}^{*}$-weighted images at TE = 70 ms for all slices of the same subjects shown in Figure 7 (i.e., Subject 3: 27.9 weeks, Subject 9: 35.6 weeks). The datasets were acquired in a slice-interleaved manner and reordered for video formatting. While S1 demonstrates the proposed radial approach can produce high-resolution $R_{2}^{*}$ maps and contrast-weighted images for a fetal brain with rapid motion, S2 shows high-resolution $R_{2}^{*}$ maps and images can be readily achieved by the proposed approach when the fetal brain remains more stable.

	\section*{Discussion}
	\label{sec:Discussion}

    In this work, we present a rapid, high-resolution, and distortion-free $R_{2}^{*}$
  mapping technique for the fetal brain. With the multi-echo radial sequence offering motion robustness and efficient k-space coverage, the regularized calibrationless model-based reconstruction efficiently estimates quantitative maps and coil sensitivity maps directly from undersampled k-space data. Validation through simulations, phantom studies, and data from nine fetal subjects confirms reliable and accurate 
$R_{2}^{*}$
  measurements compared to reference methods. The proposed approach achieves distortion-free fetal brain $R_{2}^{*}$ mapping at a nominal resolution of $1.1 \times 1.1 \times 3$ mm$^{3}$ within 2 seconds. 
  In addition, it enables the synthesis of high-resolution
$R_{2}^{*}$-weighted images, offering complementary information to the conventional T2-weighted HASTE images for fetal imaging.

To the best of our knowledge, this is the first study utilizing motion-robust 2D radial acquisition for rapid, high-resolution, and distortion-free 
$R_{2}^{*}$
  mapping of the fetal brain. Following validating the proposed approach using numerical and experimental phantoms, we compared the model-based reconstruction with the PICS with Graph Cut method on the same radial datasets. The latter represents the state-of-the-art technique for water-fat separation and quantitative $R_{2}^{*}$ mapping in body imaging applications. Our results demonstrate strong agreement between the two reconstruction methods, indicating that both are effective for fetal brain parameter quantification. Furthermore, the model-based approach provides enhanced image detail compared to the PICS with Graph Cut method, likely due to its direct reconstruction of parameter maps from k-space and the application of regularization directly to the quantitative $R_{2}^{*}$ maps \cite{Wang_Philos.Trans.R.Soc.A._2021}. Additionally, while there was motion between repeated scans, our results show high repeatability (reliability) of the generated maps by the proposed radial acquisition with model-based reconstruction.

  Compared to conventional multi-echo EPI methods, the proposed radial approach provides improved spatial resolution and reduced distortion. Quantitative analysis shows comparable $R_{2}^{*}$ values between the two acquisition strategies, with the remaining differences likely attributed to variations in fetal position, spatial resolution, and distortion. Both radial and EPI acquisitions yielded slightly higher $R_{2}^{*}$ values than those reported in the literature, particularly in the THA regions. This could be due to the age difference in the studied fetal groups as $R_{2}^{*}$ values change rapidly along the gestation age. A more detailed analysis of $R_{2}^{*}$ variation across age and between subjects warrants a larger scale study, which requires enrolling and scanning a larger number of subjects. 

  As noted in the Introduction, EPI is highly efficient due to its long readout and has been widely used in fetal imaging \cite{afacan2019fetal}. However, its long readout introduces distortion from $B_{0}$ inhomogeneity and blurring from $T_2^*$ decay. Moreover, for accurate $R_2^*$ estimation, multi-echo EPI typically requires shorter readouts, compromising its spatial resolution for temporal resolution \cite{wang2019echo}.
 In contrast, the proposed radial FLASH employs a much shorter readout (i.e., around 2 ms per spoke), making it less sensitive to $B_{0}$ inhomogeneity. Furthermore, complementary spokes from different excitations and echoes are designed and combined with time-resolved reconstruction methods (e.g., model-based reconstruction) in this study, enabling the generation of high-resolution, blurring-free, multi-contrast images and/or quantitative maps. 
Regarding acoustic noise, previous studies have shown that FLASH sequences generate moderate acoustic noise levels, whereas EPI sequences produce higher noise due to rapid gradient switching \cite{mcjury2022acoustic}. In this work, the introduction of blip gradients in the multi-echo FLASH sequence requires similar rapid gradient switching, resulting in noise levels comparable to those of EPI. 

Stack-of-stars radial multi-echo acquisitions \cite{Block_J.KoreanSoc.Magn.Reson.Med._2014} have been employed for 3D $R_{2}^{*}$
  mapping of the placenta \cite{armstrong20193d}, and our previous work extended this approach to fetal brain $R_{2}^{*}$
  mapping \cite{Wang_ISMRM_2024}. While these 3D methods perform well for fetal brains with minimal or no motion, they require extended acquisition times (typically over 3 minutes), posing challenges in cases of rapid fetal brain motion, even with advanced motion correction techniques. In contrast, the proposed 2D technique delivers reliable 
$R_{2}^{*}$  maps within a short acquisition window, demonstrating robustness in scenarios with significant fetal motion. Moreover, the proposed method is very general and can be extended to the quantification of other challenging fetal organs. For instance, high-resolution quantitative $R_{2}^{*}$ mapping of the fetal liver is of great interest as it could provide valuable insights into evaluating liver iron overload in the fetal stage.




As a technical development study, this work is limited by the relatively small sample size. The limited number of subjects prevented an identification of clear trends in fetal brain $R_{2}^{*}$ with respect to gestational age. Future studies will apply the technique to a larger cohort to investigate brain development, focusing on how $R_{2}^{*}$ values evolve across gestational age. Moreover, the current approach may not be able to provide $R_{2}^{*}$ maps with coherent anatomic boundaries in 3D because of 1) relatively thick slices, and 2) inter-slice motion.  Ongoing efforts are focused on acquiring multi-orientation 2D data and applying motion-corrected slice-to-volume reconstruction to generate high-resolution 3D volumes.  However, this is not a trivial task and requires additional development of reliable slice-level motion correction and image reconstruction methods \cite{gholipour2010robust, uus2020deformable}. 

 Furthermore, the 2-second acquisition time, while effective, remains longer than the HASTE sequence. The latter typically takes less than 1 second and is highly effective at freezing motion. Consequently, although the proposed radial acquisition is robust to motion, this method may still be 
 affected by very rapid motion during data acquisition. In cases of significant fetal motion during acquisition, two complementary strategies could be exploited. First, spoke-wise motion detection methods could be developed to identify and either exclude or correct motion-corrupted radial spokes. Second, further reducing acquisition time will be critical to mitigate the impact of rapid motion. 
Our retrospective analysis of Subject 9 (Supporting Information Figure S5) demonstrates that the proposed method can produce reasonable images within 1 second, albeit with increased noise and reduced $R_{2}^{*}$ accuracy. 
To improve this, future work will focus on further shorten acquisition time without compromising accuracy or precision. One direction involves replacing the hand-crafted $\ell_{1}$-Wavelet transform with a deep-learning-enhanced regularizer \cite{blumenthal2024self, li2022accelerating, li2024accelerated} in the model-based reconstruction. Another idea would be to adapt radial simultaneous multi-slice techniques \cite{Wang_Magn.Reson.Med._2020} for sub-second quantitative fetal brain imaging.

	\section*{Conclusion}
	\label{sec:Conclusion}
This work demonstrates the feasibility of radial acquisition for motion-robust quantitative $R_{2}^{*}$ mapping of the fetal brain. By combining multi-echo radial FLASH with calibrationless model-based reconstruction, the proposed method achieves accurate, distortion-free fetal brain  $R_{2}^{*}$  mapping at a nominal resolution of $1.1 \times 1.1 \times 3$ mm$^{3}$ within 2 seconds.
	
\section*{Conflict of Interest}

Dr. Hongli Fan is an employee of Siemens.

\section*{Open Research}
\subsection*{Data Availability Statement}
In the spirit of reproducible research, code and data to
reproduce the reconstruction and analysis in this work will be available on
\url{https://github.com/IntelligentImaging/FetalR2Star}.


	\bibliographystyle{mrm}
	\bibliography{radiology}
	
	\clearpage

Supporting Information
         


\begin{figure}[H]
	\centering
	\includegraphics[width=0.95\textwidth]{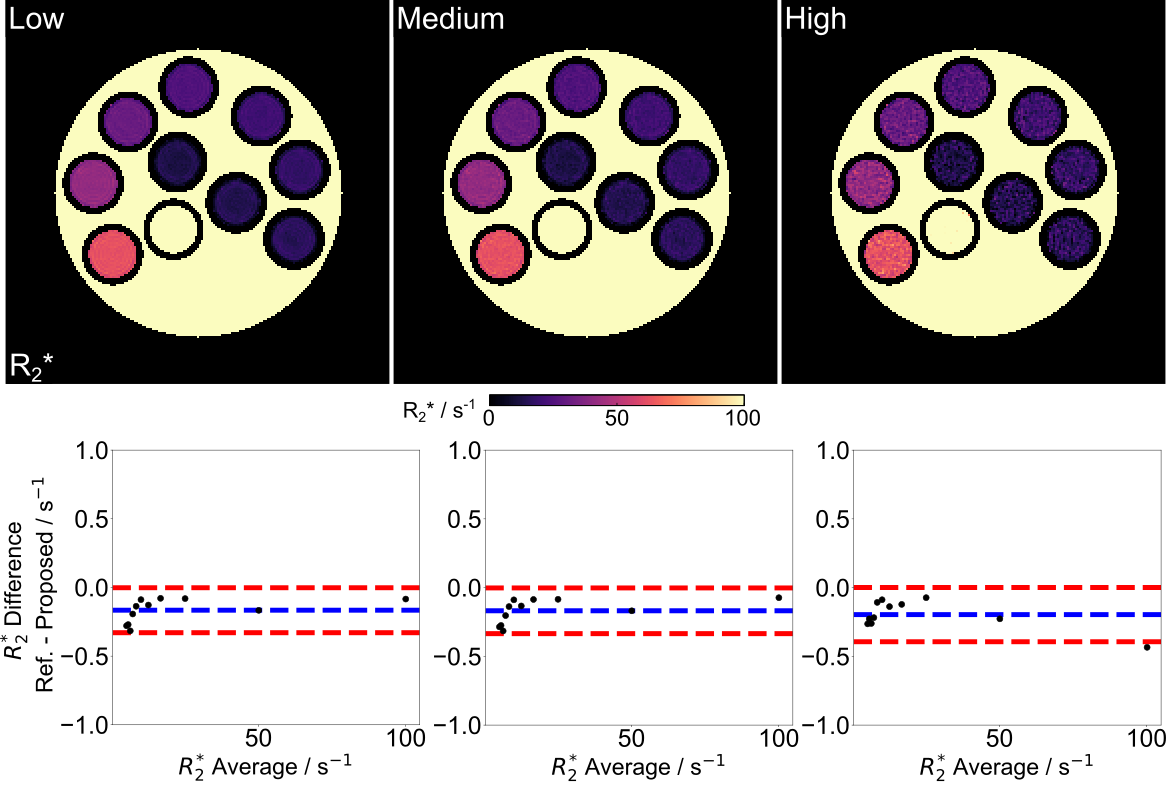}
	\caption*{Supporting Information Figure S2. (Top) Quantitative $R_{2}^{*}$ maps estimated from simulated data using the proposed model-based approach under (left) low, (middle) medium, and (right) high levels of Gaussian noise.
		(Bottom) Bland–Altman plots comparing ROI-based $R_{2}^{*}$ estimates between the proposed method and ground truth. The mean differences are $-0.17 \pm 0.06$ $s^{-1}$, $-0.17 \pm 0.07$ $s^{-1}$, and $-0.18 \pm 0.07$ $s^{-1}$, respectively. Detailed quantitative values for each tube (mean ± standard deviation) are provided in Supporting Information Table S1.} 
\end{figure}

\begin{table} 
	\centering
	\caption{\normalsize Supporting Information Table S1. Quantitative $R_{2}^{*}$ values ($s^{-1}$, mean $\pm$ SD) for the numerical phantom with different noise levels in the Supporting Information Figure S2. }
		\vspace{-10pt}
		\begin{center}
			\begin{tabular}{c| c| c| c| c}
				\hline
				\multirow{2}{*}{True $T_{2}^{*}$ / ms}& \multirow{2}{*}{True $R_{2}^{*}$ / $s^{-1}$} & \multicolumn{3}{c}{Estimated $R_{2}^{*}$ / $s^{-1}$ across noise levels}\\ \cline{3-5}
				
				& & Low & Medium  &High \\
				
				\hline
				10  &	100  &$100.2 \pm 2.1$   &$100.2 \pm 2.3$  &  $100.2 \pm 7.3$    \\
				20  &	 50    & $50.1 \pm 0.9$  &  $50.1 \pm 1.4$  & $50.2 \pm 6.2$   \\
				40  &	25   & $25.1 \pm 0.6$ & $25.1 \pm 0.8$  & $25.1 \pm 2.6$  \\
				60  &	16.7  &  $16.7 \pm 0.6$  & $16.8 \pm 0.9$ & $16.8 \pm 2.7$   \\
				80  &	12.5     &  $12.6 \pm 0.6$ & $12.6 \pm 0.8$  & $12.6 \pm 2.5$  \\
				100  & 10     &  $10.1 \pm 0.6$  &  $10.1 \pm 0.8$  & $10.1 \pm 2.2$  \\
				120  &8.3  &  $8.5 \pm 0.6$  &$8.5 \pm 0.8$  & $8.4 \pm 2.4$ \\
				140  &7.1    &  $7.3 \pm 0.6$ &$7.3 \pm 0.9$   & $7.4 \pm 2.5$ \ \\
				160  &6.3    &  $6.6 \pm 0.7$  &$6.6 \pm 0.9$  & $6.5 \pm 2.7$ \\
				180  &5.6    &  $5.8 \pm 0.8$&$5.8 \pm 0.9$  & $5.8 \pm 2.9$ \\
				200  &5.0   &  $5.2 \pm 0.7$  &$5.2 \pm 0.9$   & $5.3 \pm 2.5$ \\

				\hline
				
				\hline
			\end{tabular}\\
		\end{center}
	
	\label{tab:}
	\vspace{-20pt}
\end{table}

 \begin{figure}[H]
	\centering
	\includegraphics[width=1.0\textwidth]{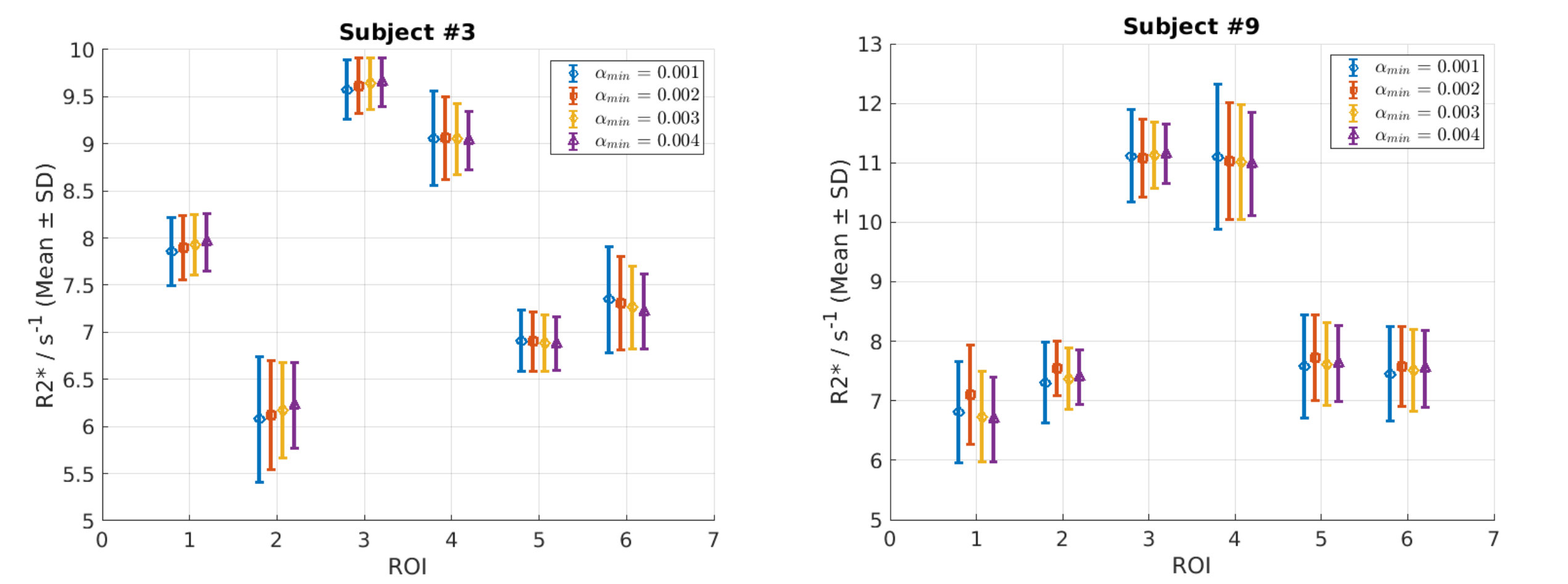}
	\caption*{Supporting Information Figure S3. Quantitative $R_{2}^{*}$ values (mean and standard deviation) within ROIs that were manually drawn into the frontal white matter, thalamic gray matter, and occipital white matter regions of all $R_{2}^{*}$ maps in Figure 3.} 
\end{figure}

\begin{figure}[H]
	\centering
	\includegraphics[width=1.0\textwidth]{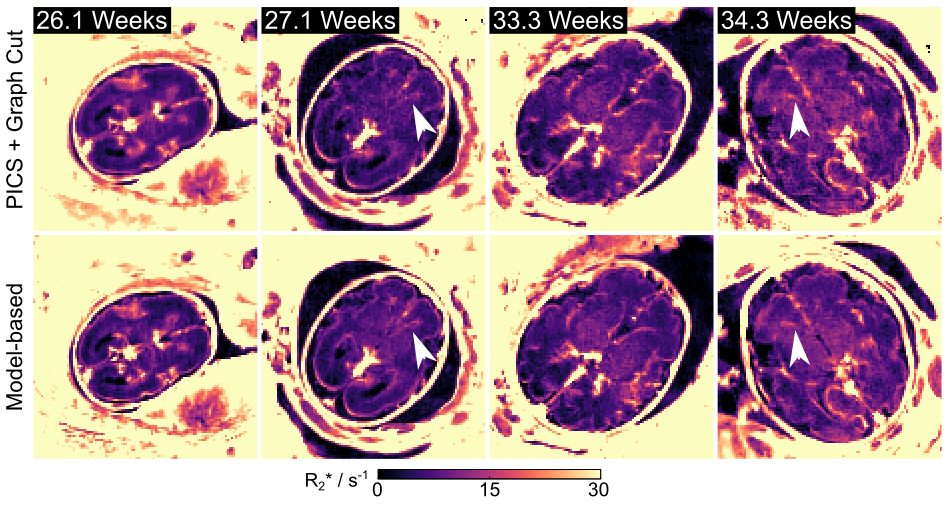}
	\caption*{Supporting Information Figure S4. Comparison of quantitative fetal brain $R_{2}^{*}$ maps estimated using (top) PICS with Graph Cut and (bottom) model-based reconstruction for the remaining four subjects at different gestational ages. Similar to Figure 5 (A), white arrows indicate improved image details by model-based reconstruction. Quantitative comparison of all subjects is presented in Figure 5 (B).}
\end{figure}

\begin{figure}[H]
	\centering
	\includegraphics[width=1.0\textwidth]{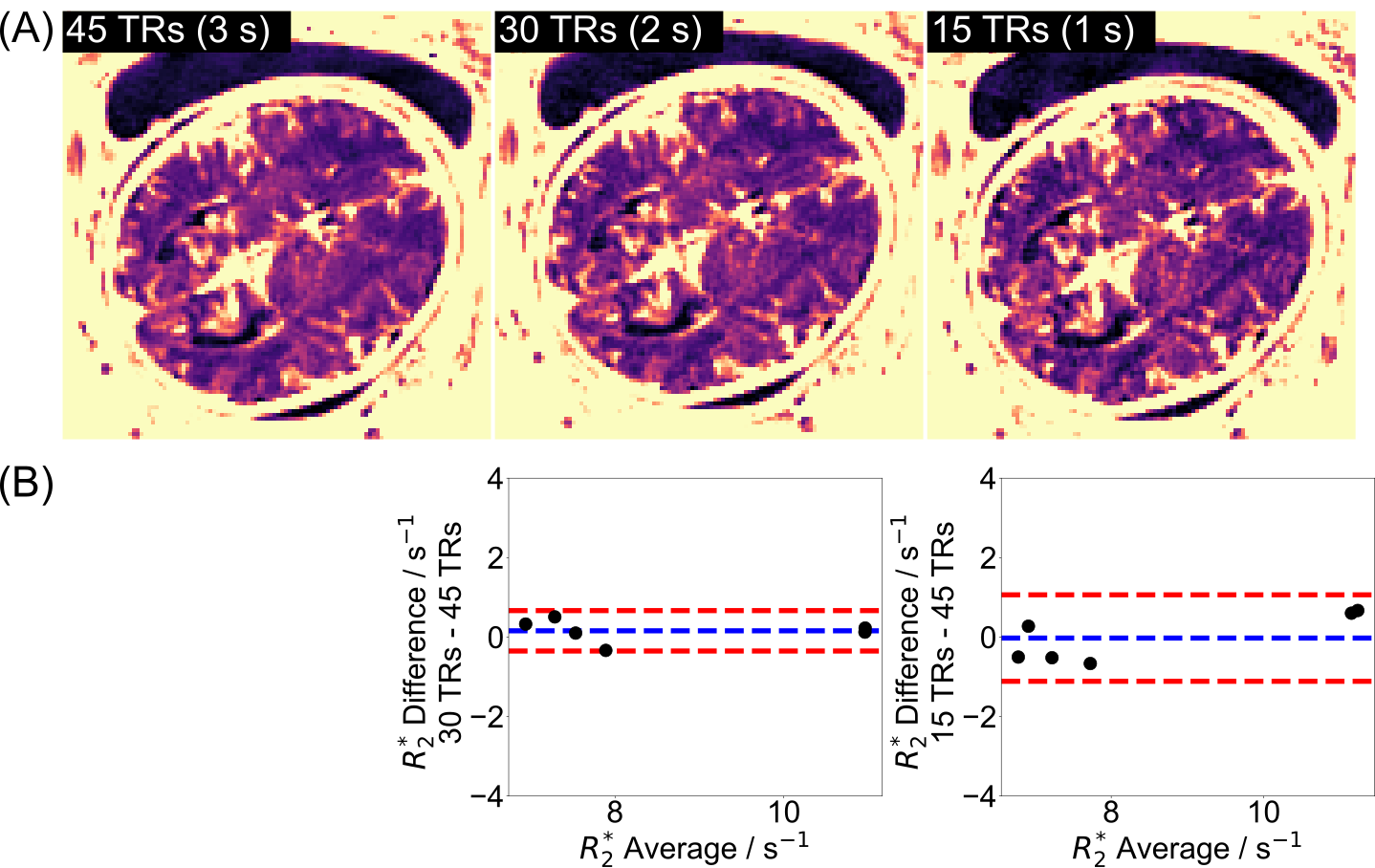}
	\caption*{Supporting Information Figure S5. (A). Quantitative $R_{2}^{*}$ maps estimated using 3-second (45 TRs), 2-second (30 TRs), and 1-second (15 TRs) multi-echo radial FLASH acquisitions for Subject 9. Because of minimal motion during data acquisition, the 2-second and 1-second data were retrospectively undersampled from the 3-second one. (B) Bland-Altman plots comparing mean $R_{2}^{*}$ values between the 3-second and 2-second, and 3-second and 1-second acquisitions. The mean $R_{2}^{*}$ differences are $0.16 \pm 0.26$ s$^{-1}$ and $-0.02 \pm 0.6$ s$^{-1}$, respectively.}
\end{figure}

\textbf{Supporting Information Video S1.}  Quantitative $R_{2}^{*}$ maps and synthesized $R_{2}^{*}$-weighted images (TE = 70 ms) for 16 slices of Subject 3 (27.9 weeks).

\textbf{Supporting Information Video S2.}  Quantitative $R_{2}^{*}$ maps and synthesized $R_{2}^{*}$-weighted images (TE = 70 ms) for 20 slices of Subject 9 (35.6 weeks).

 \begin{table} 
	\centering
	\caption{\normalsize Quantitative $R_{2}^{*}$ values (s$^{-1}$, mean $\pm$ SD) for fetal brains. }
	\vspace{-10pt}
	\begin{center}
		\begin{tabular}{c| c| c| c}
			\hline
{Tissue} & FWM & THA  &OWM \\
			\hline
Radial 3T (Model-based) &	$6.1 \pm 1.0$ & $9.1 \pm 1.3$ &$6.0 \pm 1.3$     \\
Radial 3T (PICS + Graph Cut)  &	$6.2 \pm 1.0$ & $9.2 \pm 1.4$ &$5.9 \pm 1.5$     \\
EPI	3T &	 $5.7 \pm 1.1$  & $8.7 \pm 1.1$ &  $6.4 \pm 1.6$  \\
Rivkin et al.\cite{rivkin2004prolonged} 1.5 T &	6.6   &  7.9  &  \\
Vasylechko et al.\cite{vasylechko2015t2} 1.5 T &	4.3   &  6.5   & 4.0  \\
Blazejewska et al.\cite{blazejewska20173d} 1.5 T &	3.9    &  6.0 &  \\
Vasylechko et al.\cite{vasylechkofetal3T} 3.0 T &	5.0   &  6.6  & 4.4 \\

		\hline

		\hline
		\end{tabular}\\
	\end{center}
	
	\label{tab:example}
		\vspace{-20pt}
\end{table}

\newpage

\newpage

	\section*{Acknowledgements}This work was supported by National Institutes of Health (NIH) under award numbers R01NS106030, R01EB031849, R01EB032366, R01HD109395, R01EB032708, R01LM013608, R01EB019483, R01NS133228, R01NS121657 and U24EB029240; in part by the Office of the Director of the NIH under award number S10OD025111, and in part by NVIDIA Corporation. We are grateful to Dr. Ellen Grant for the insightful comments and Dr. Borjan Gagoski for sharing the NIST phantom. 
\end{document}